\documentclass[sigconf]{acmart}
\acmConference[ICSE 2022]{The 44th International Conference on Software Engineering}{May 21–29, 2022}{Pittsburgh, PA, USA}

\AtBeginDocument{%
  \providecommand\BibTeX{{%
    \normalfont B\kern-0.5em{\scshape i\kern-0.25em b}\kern-0.8em\TeX}}}

\usepackage[utf8]{inputenc}

\usepackage{amssymb,amsfonts}
\usepackage{textcomp}
\usepackage{xcolor}
\usepackage{hyperref}
\usepackage{microtype}
\usepackage{subcaption}
\usepackage[linesnumbered,ruled,vlined]{algorithm2e}

\usepackage{amsthm}
\theoremstyle{definition}
\newtheorem{definition}{Definition}

\usepackage{balance}

\copyrightyear{2022}
\acmYear{2022}
\setcopyright{acmcopyright}\acmConference[ICSE '22]{44th International Conference
on Software Engineering}{May 21--29, 2022}{Pittsburgh, PA, USA}
\acmBooktitle{44th International Conference on Software Engineering (ICSE '22), May
21--29, 2022, Pittsburgh, PA, USA}
\acmPrice{15.00}
\acmDOI{10.1145/3510003.3510078}
\acmISBN{978-1-4503-9221-1/22/05}

\begin{document}

\title{Conflict-aware Inference of Python Compatible Runtime Environments with Domain Knowledge Graph}


\author{Wei Cheng}
\affiliation[obeypunctuation=true]{
	\department{State Key Laboratory for Novel Software Technology}\\
	\institution{Nanjing University},
	\country{China}}
\email{wchengcs.nju@gmail.com}

\author{Xiangrong Zhu}
\affiliation[obeypunctuation=true]{
	\department{State Key Laboratory for Novel Software Technology}\\
	\institution{Nanjing University},
	\country{China}}
\email{xrzhu.nju@gmail.com}

\author{Wei Hu}
\authornote{Corresponding author}
\affiliation[obeypunctuation=true]{
	\department{State Key Laboratory for Novel Software Technology}\\
	\institution{Nanjing University},
	\country{China}}
\email{whu@nju.edu.cn}




\begin{abstract}
Code sharing and reuse is a widespread use practice in software engineering. 
Although a vast amount of open-source Python code is accessible on many online platforms, programmers often find it difficult to restore a successful runtime environment. 
Previous studies validated automatic inference of Python dependencies using pre-built knowledge bases. 
However, these studies do not cover sufficient knowledge to accurately match the Python code and also ignore the potential conflicts between their inferred dependencies, thus resulting in a low success rate of inference. 
In this paper, we propose PyCRE, a new approach to automatically inferring Python compatible runtime environments with domain knowledge graph (KG). 
Specifically, we design a domain-specific ontology for Python third-party packages and construct KGs for over 10,000 popular packages in Python 2 and Python 3. 
PyCRE discovers candidate libraries by measuring the matching degree between the known libraries and the third-party resources used in target code. 
For the NP-complete problem of dependency solving, we propose a heuristic graph traversal algorithm to efficiently guarantee the compatibility between packages. 
PyCRE achieves superior performance on a real-world dataset and efficiently resolves nearly half more import errors than previous methods.
\end{abstract}



\maketitle

\begin{figure*}
    \centering
    \includegraphics[width=\textwidth]{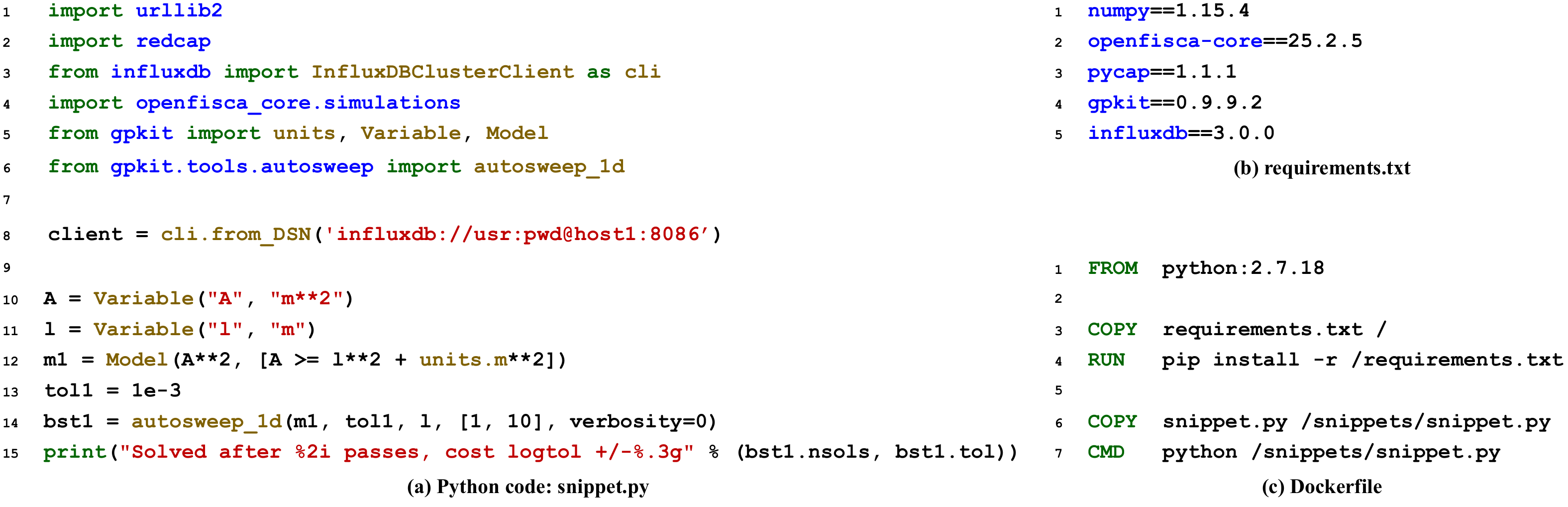}
    \caption{A motivating example.}
    \label{fig:fig_example}
\end{figure*}

\section{Introduction}
\label{sec:intro}

With the rise of programming communities such as GitHub and StackOverflow, code sharing and reuse have become a common practice for programmers \cite{yang2017stack}.
Python is one of the most popular high-level programming languages today, due in part to its massive third-party package resources, which also often cause environment configuration issues. 
An empirical research \cite{yang2016query} observed that the usability rate of all the Python code snippets on StackOverflow is 76\% parsable and 25\% runnable, and the work in \cite{horton2018gistable} further found that 75.6\% of the Python code snippets shared through GitHub are not executable with over half of failures due to missing dependencies in a clean Python 2 environment.

Executing a Python code that contains third-party resources in a clean Python environment triggers dependency errors, which raise the built-in exception \texttt{ImportError}.
To resolve this, programmers need to specify the packages and desired versions in a configuration script such as \textit{requirements.txt}. 
However, this is not a trivial work. 
The study in \cite{horton2018gistable} observed that programmers usually spend between 20 minutes and two hours to set up the environment, and in some cases, they even cannot restore a correct execution environment. 
Therefore, automatically inferring the runtime environments of Python code helps, which can free up the time of programmers spent on dependency issues, and thus is significant for code reuse and automated software configuration management. 

However, there are several challenges to automatic inference of Python runtime environments. 
Let us see a Python code snippet shown in Figure~\ref{fig:fig_example}a.
The imported top-level modules, namely \textit{redcap}, \textit{influxdb}, \textit{openfisca\_core} and \textit{gpkit}, are not built-in modules in the Python standard library.
Executing \texttt{pip install redcap} receives the error message \texttt{ERROR: No matching distribution found for redcap}, because \textit{redcap} does not exist on Python Package Index (PyPI) and the package corresponding to it is \textit{pycap}. 
In fact, it is common in practice that the name of a module imported in Python code does not match the name of the Python package it belongs to. 
Moreover, after a successful installation with \texttt{pip install influxdb}, another error message \texttt{ImportError: cannot import name InfluxDBClusterClient} appears, which indicates that the latest version \textit{influxdb-5.3.1} does not contain this attribute. In fact, the last version containing attribute \textit{influxdb.InfluxDBClusterClient} is \textit{3.0.0}.

A naive approach installs the Python packages with the same names as the imported top-level modules. 
However, this approach fails to infer correct dependencies in many cases, as illustrated in the above example. 
The challenges are essentially the lack of sufficient domain knowledge.
DockerizeMe \cite{horton2019dockerizeme} builds an offline knowledge base semi-automatically to infer the environment dependencies for Python code snippets.
V2 \cite{horton2019v2} searches working environments with proper versions based on the error messages of code execution.
SnifferDog \cite{wang2021restoring} builds an API bank of Python packages to infer the specific versions and restore the execution environments of Jupyter notebooks.
These previous studies have achieved good performance by using pre-built knowledge bases, but they do not cover sufficient knowledge to accurately match more complex Python code. 
In this paper, we design an elaborated ontology for Python third-party packages and automatically construct Python package knowledge graphs (KGs) by installing and analyzing the releases on PyPI. 

After building the domain KGs, we are still challenged by how to match the target code with Python dependencies. 
To cope with this, we design a novel metric of matching degree and treat all attributes under a top-level module as a whole to better discover required libraries.
Furthermore, 
dependency conflicts occur when different inferred packages depend on the same package, but specify different and incompatible versions of that package. 
Continuing to consider the code shown in Figure~\ref{fig:fig_example}a, installing the latest versions of packages \textit{openfisca-core} and \textit{gpkit} causes a dependency conflict, because \textit{openfisca-core-25.2.5} requires \textit{numpy$<$1.16,$\ge$1.11} and \textit{gpkit-0.9.9.9.1} requires \textit{numpy$\ge$1.16.4}. 
Thus, we should choose an older version \textit{gpkit-0.9.9.2}, which requires \textit{numpy$\ge$1.13.3} and is compatible with other packages. 
Dependency solving should ensure that all direct dependencies and transitive dependencies (i.e. dependencies of dependencies) are compatible with the rest of inferred environment. 
To the best of our knowledge, all previous studies \cite{horton2019dockerizeme, horton2019v2, wang2021restoring} have not considered the compatibility of the inferred Python environments yet. 
Due to the NP-completeness of dependency solving \cite{mancinelli2006managing}, we propose a heuristic graph traversal algorithm to infer a compatible environment, which efficiently selects the newer versions and prunes the traversal paths.


The main contributions of this paper are listed as follows:
\begin{itemize}
    \item We design an ontology for Python third-party packages and an automatic approach to KG construction.
    As a result, we create the Python package KGs for Python 2 and Python 3, each of which contains the knowledge of over 10,000 Python packages and nearly 300,000 versions. (Sections~\ref{sec:kg})
    \item We define a new metric of matching degree between Python libraries and third-party resources in target code to discover required libraries.
    Moreover, we consider the compatibility of the inferred Python environments, and design an efficient heuristic algorithm for dependency solving. (Sections~\ref{sec:alg})
    \item We implement our approach called PyCRE (\url{https://github.com/nju-websoft/PyCRE}) and evaluate it with 10,250 real-world Python code snippets on Gistable~\cite{horton2018gistable}. 
    Our experiments show that PyCRE efficiently resolves dependency issues for both Python 2 and Python 3, leaving only 1,524 \texttt{ImportError}, which is significantly superior to 2,654 \texttt{ImportError} of the state-of-the-art approach \cite{horton2019v2}.
\end{itemize}


\section{Overview of PyCRE} 
\label{sec:pycre}

\begin{figure}[tb]
  \centering
  \includegraphics[width=\linewidth]{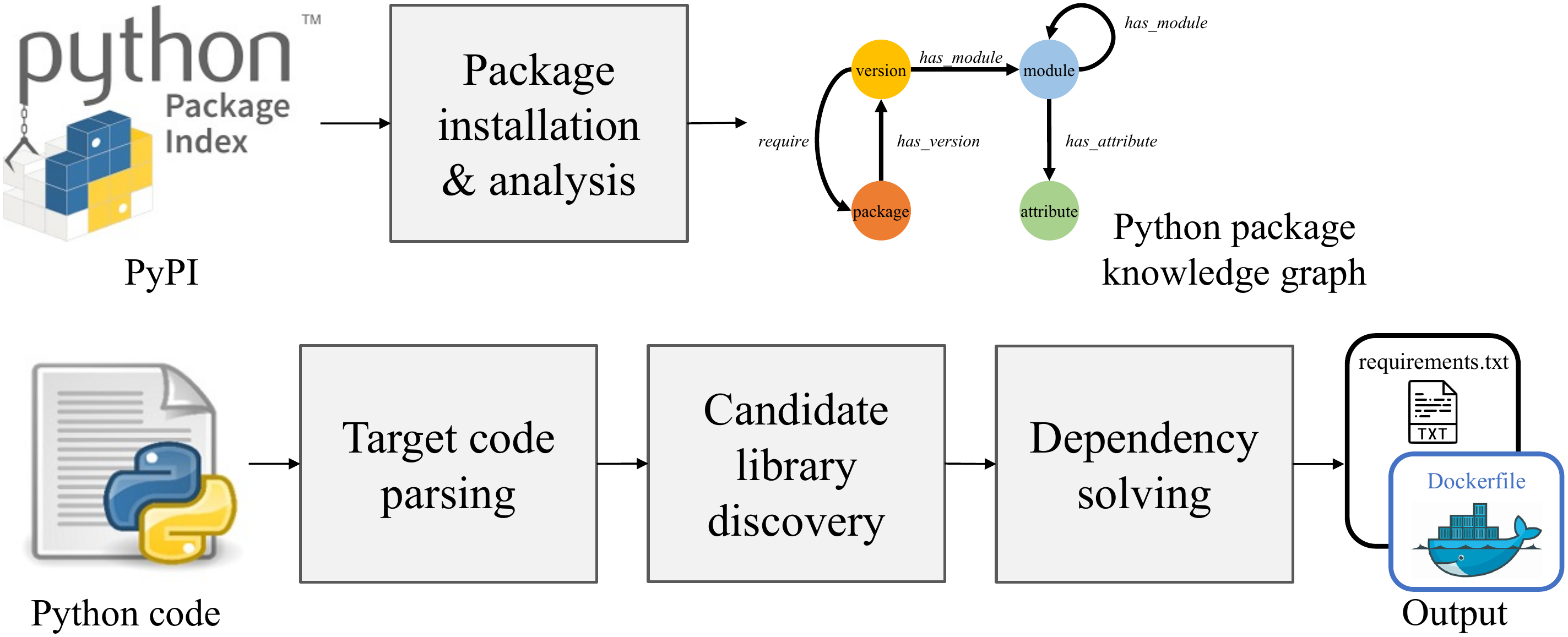}
  \caption{Overview of our approach.}
  \label{fig_overview}
\end{figure}

Figure~\ref{fig_overview} depicts an overview of PyCRE, which automatically infers Python compatible runtime environments with the pre-built Python package KGs. 
The fundamental requirement of PyCRE is to cover as many third-party resources used in the target code as possible.
Additionally, PyCRE directly specifies the packages and their versions in a feasible installation order to avoid the potential conflicts between inferred dependencies. 

PyCRE consists of two phases, where the upper portion of the figure shows the construction of our Python package KGs. 
All we can know from target code are the imported modules and the called attributes, including the names of variables, classes, functions and even hidden submodules. 
See Figure~\ref{fig:fig_example}a for example, matching the third-party resources in target code to the correct libraries requires a great deal of domain knowledge. 
According to our designed ontology, we offline construct two Python package KGs for Python 2 and Python 3, respectively.

The bottom portion shows the automatic inference of compatible runtime environments.
With the Python package KGs, PyCRE parses a target Python code, discovers its candidate libraries according to matching degrees, and generates the compatible runtime environment through dependency solving. 

The output of PyCRE is a \textit{requirements.txt} containing a list of required Python packages with specific versions in a correct order and a \textit{Dockefile} containing the inferred Python version. 
The dependencies in \textit{requirements.txt} work together, reducing the risk of dependency conflicts compared to installing dependencies individually.
For example, Figure~\ref{fig:fig_example}b shows the \textit{requirements.txt} generated by PyCRE for the Python code shown in Figure~\ref{fig:fig_example}a. 
The \textit{Dockerfile} shown in Figure~\ref{fig:fig_example}c installs all Python packages with the command \texttt{pip install -r requirements.txt}.

\section{Python Package Knowledge Graph} 
\label{sec:kg}

Domain knowledge of Python packages is essential for automatic inference of compatible runtime environments. 
We devote our efforts to designing an ontology for Python third-party packages and a method to automatically construct the corresponding KGs. 

\subsection{Python Package Ontology Design}
\begin{figure}[tb]
  \centering
  \includegraphics[width=0.85\linewidth]{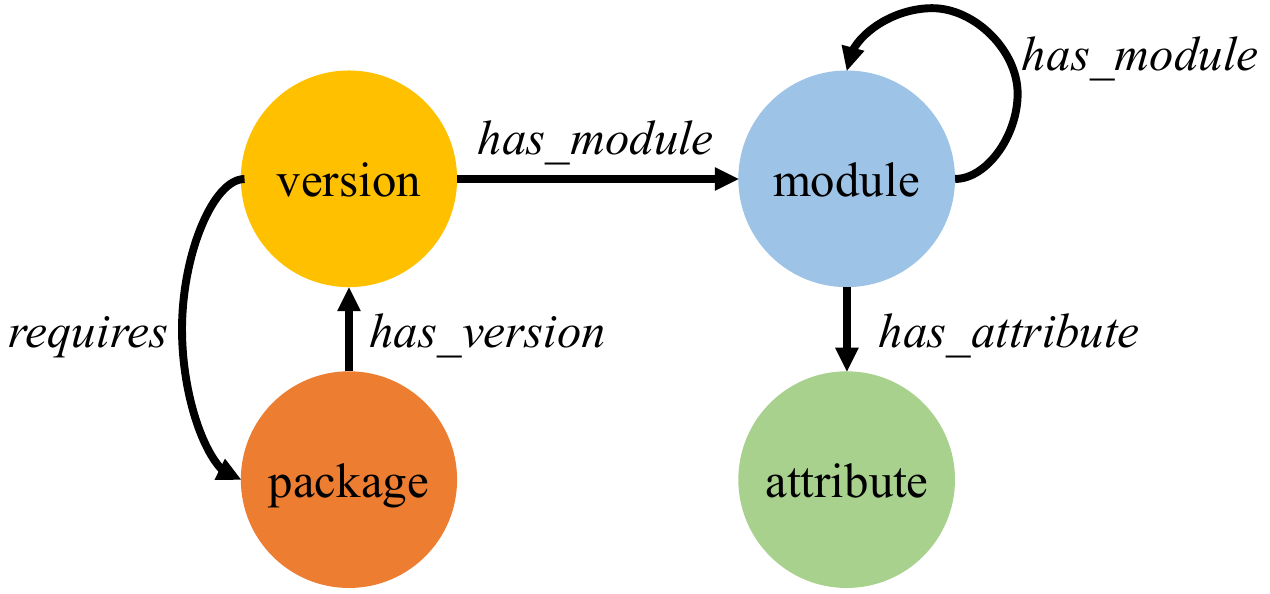}
  \caption{Entities and relationships defined in the Python package ontology.}
  \label{fig_kg}
\end{figure}

As shown in Figure~\ref{fig_kg}, we define an ontology to represent relationships between entities and properties for describing entities:
\begin{itemize}
    \item \textbf{Package node.} Each package node represents a Python package and stores the package's \textit{name} as a property. 
    The stored package names are normalized and unified, making no two package entities having an identical name.
    \item \textbf{Version node.} Each version of a Python package is stored as a distinct version node. 
    A version node contains its standard \textit{version} identifier and \textit{install\_status} of the corresponding release. 
    There are three values for \textit{install\_status}: \textit{Success}, \textit{Fail} and \textit{Unknown}, where \textit{Unknown} means that the version has not been installed yet.
    \item \textbf{Module node.} Each module node corresponds to a specific module of a version and has two properties. 
    Property \textit{import\_status} takes the value \textit{True} or \textit{False}, indicating whether the module can be successfully imported or not. 
    Another property is the fully-qualified \textit{name}, e.g., \textit{client} is a submodule of module \textit{redis}, and its name is stored as \textit{redis.client}. 
    However, module names are not unique, as different versions of a package may contain homonymous modules that have different attributes, or even different packages may have homonymous modules.
    \item \textbf{Attribute node.} Each attribute node stores the attribute's \textit{name} as a property. 
    Unlike module nodes, attributes with an identical name are defined as a single entity in the ontology. 
    For example, attribute \textit{redis.client.Redis} is saved as attribute \textit{Redis} of module \textit{redis.client}. 
    In our ontology, attribute can be variable, function, class or any content available in the corresponding module.
    \item[$\circ$] \textbf{Package $\to$ Version:} \textit{has\_version} edge, which indicates the relationship between the package and its version.
    \item[$\circ$] \textbf{Version/Module $\to$ Module:} \textit{has\_module} edge. Each successfully installed version has modules in principle, and modules may have their submodules.
    \item[$\circ$] \textbf{Module $\to$ Attribute:} \textit{has\_attribute} edge, which shows that the attribute is available in the module.
    \item[$\circ$] \textbf{Version $\to$ Package:} \textit{requires} edge, which represents that the package is a direct dependency of the version, and has a \textit{requirement} property to store the version specifier. 
\end{itemize}

\subsection{Knowledge Acquisition} 
\label{sec:ka}
To extract knowledge, we first determine the Python packages that we want to analyze and get all available versions. 
Then, we record the installation status by installing each version of the packages. 
At last, we acquire the modules, attributes and direct dependencies of those versions that are successfully installed. 
The whole process of knowledge acquisition is automated.

\begin{itemize}
\item\textbf{Packages.}
A list of available distributions on PyPI can be obtained according to PyPISimple.\footnote{\url{https://wiki.python.org/moin/PyPISimple}}
Due to the expense of time and storage, we usually specify a list of packages for further knowledge acquisition.
With the knowledge of commonly used Python packages in a certain domain, the success rate of inferred Python compatible runtime environments can be greatly improved. 
\item\textbf{Versions.}{
We get all available versions of each package by executing \texttt{pip install $<$package$>==<$version$^*$$>$}, since many packages are not available in exactly the same versions under different Python releases.
\texttt{$<$version$^*$$>$} is a special version identifier that does not exist, which makes pip output all available versions.
}
\item\textbf{Installations.}{
Some releases have strict requirements for the supported Python versions or even require system-level dependencies that pip cannot handle. 
Thus, not every release on PyPI can be installed successfully. 
We attempt to install each version of a package with \texttt{pip install $<$package$>==$} \texttt{$<$version$>$} and record the installation status.
}
\item\textbf{Requirements.}{
Most releases require certain direct dependencies to be installed before their installations, which are stored in their metadata. 
We get the dependency requirements of each successfully installed release from its \textit{METADATA} file. 
Dependency requirements prompt a lot of unknown Python packages, due to the incompleteness of our KG. We create Package nodes for these packages, but without doing further knowledge acquisition.
}
\item\textbf{Modules and attributes.}{
A Python distribution usually has multiple modules. 
For each successfully installed distribution, we first attempt to get its top-level modules from \textit{top\_level.txt}. 
If that file does not exist, we try to check all created directories and files to obtain its top-level modules. 
We then find all submodules of each top-level module recursively. 
Finally, we try to import each module and use Python built-in function \textit{dir()} to attain all attributes of the imported modules. 
We remove the modules and attributes that start with the underscore because they are conventionally intended for internal use.
}

\end{itemize}

After constructing the Python package KGs, we can periodically check for new packages and versions based on the above process to incrementally upgrade our KGs. 
Note that PyPI doesn’t support the replacement of existing releases but only deletion (we can also delete the corresponding entities and relations in our KGs), thus ensuring the consistency of our KGs. 

\section{Environment Inference}
\label{sec:alg}

For environment inference, we first obtain the imported third-party modules and called attributes by parsing the target code. 
Then, we query our KGs to discover candidate libraries that best match these modules and attributes. 
Finally, we expand a dependency graph with transitive dependencies of candidate libraries, and infer the installation instructions of compatible dependencies in order by dependency solving.

\subsection{Target Code Parsing} 
\label{sec_parse}

As the only input, we assume that the code should be fully parsed. 
To determine which third-party libraries to be installed, we find all the imported modules that are not in the Python standard library and the called attributes in those modules. 
We first parse the target code into an abstract syntax tree (AST) in a clean Python environment, and then walk in the AST for both of the following information:
\begin{itemize}
    \item \textbf{Imported modules}. 
    The syntax for the \texttt{import} statement in Python is \texttt{import $<$module$>$ as alias} or \texttt{from $<$module$>$ import $<$name$>$ as alias}. 
    The modules in the Python standard library are ignored, and we store all possible third-party modules as \textit{imported modules}.
    It is worth mentioning that \texttt{$<$name$>$} can be a submodule, function, class, variable or any attribute that can be accessed in the module. 
    Additionally, programmers can bind an imported resource with an optional alias name and use the alias name directly in the code, but at the same time, the original name of the imported resource would no longer be used. 
    We record the mappings between the imported resources and their alias names. 
    
    \item \textbf{Called attributes}. In addition to the imported modules, the attributes of those modules used in the code are also crucial for discovering the candidate libraries. 
    We visit each attribute node in the AST and record all the attribute names prefixed with an imported resource or its alias. 
    According to the mapping obtained from the \texttt{import} statement, we then map the prefix back to the corresponding imported resource name, thus restoring its fully qualified name for each attribute. 
    For example, the fully qualified name of attribute \textit{from\_DSN} is \textit{influxdb.InfluxDBClusterClient.from\_DSN} (Line 8 in Figure~\ref{fig:fig_example}a).
    Due to the uncertainty in \texttt{from $<$module$>$ import $<$name$>$}, we treat \texttt{$<$module$>$.$<$name$>$} as a possible attribute name as well. We store these attribute names as \textit{called attributes}.
\end{itemize}

\begin{figure}[tb]
    \centering
    \includegraphics[width=\linewidth]{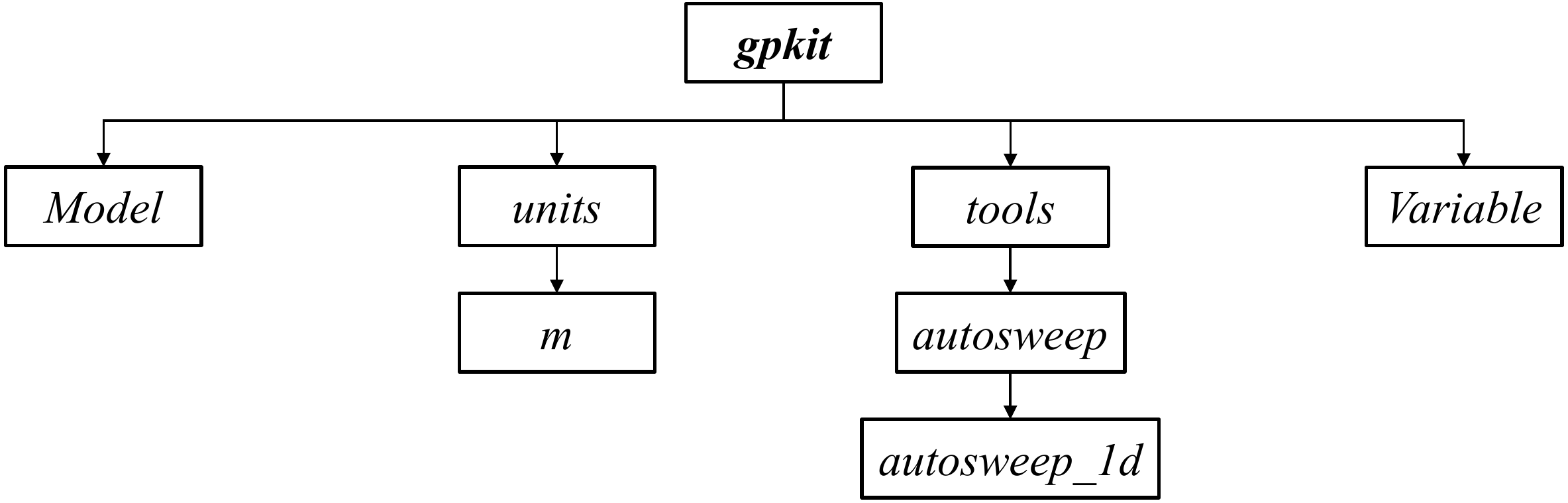}
    \caption{A parse tree generated by target code parsing.}
    \label{fig_parse}
\end{figure}

We represent the \textit{imported modules} and \textit{called attributes} as a forest. 
Each tree in the forest has a top-level module name as a root node, and contains all the submodules and attributes of that module in the code. 
For instance, by parsing the example Python code in Figure~\ref{fig:fig_example}a,  the parse tree of module \textit{gpkit} is shown in Figure~\ref{fig_parse}.

The results of target code parsing can be used to infer the candidate Python versions. 
We parse the code in Python 2 and 3, respectively, and the Python versions with the fewest \textit{imported modules} and no syntax errors are the candidates. 
For example, the imported module \textit{urllib2} (Line 1 in Figure~\ref{fig:fig_example}a) is a standard library in Python 2 and has been split into several modules in Python 3 named \textit{urllib.request} and \textit{urllib.error}. 
The sample code imports fewer non-standard modules in Python 2 than in Python 3, and thus its Python version is inferred.
If the number of \textit{imported modules} is equal, we would extrapolate further in the following.

\subsection{Candidate Library Discovery} 
\label{sec_filter}

In this phase, for each candidate Python version, we query the corresponding Python package KG to find the candidate libraries that best match the forest obtained by code parsing. 

For the resources (i.e. modules or attributes) used in the code, it is more reasonable to use partial matching than to precisely query the fully qualified names in the KG due to incomplete knowledge. 
However, widely-used string similarity metrics such as edit distance and Jaccard coefficient \cite{CohenRF03} are not applicable here, as a fully qualified name represents a reference path, e.g., \textit{openfisca\_core.simulations} and \textit{sapphire.simulations} are two completely unrelated modules. We define a metric based on the longest prefix match of the reference path to calculate the matching degree between a list of resources and a library, as shown in Algorithm~\ref{alg:match}. 
Reference paths in Python are separated by dots.
For example, the length of the reference path for resource \textit{influxdb.InfluxDBClusterClient.from\_DSN} is 3. 
Although the release \textit{influxdb-5.3.1} has attribute \textit{influxdb.InfluxDBClient}, its longest prefix reference path is \textit{influxdb} and matching degree is $\frac{1}{3}$. 

\begin{algorithm}[!t]
    \caption{$CalculatesMatchingDegree(S,L)$}
    \label{alg:match}
    \KwIn{set $S$ of resources, list $L$ of resources in the library}
    \KwOut{matching degree between $S$ and $L$}
    $degree \leftarrow 0$\;
    \ForEach{$res \in S$}
    {
        $pre \leftarrow$ longest prefix reference path of $res$ matched in $L$\;
        $degree \leftarrow degree + \frac{PathLength(pre)}{PathLength(res)}$\;
    }
    \Return $degree$\;
\end{algorithm}

We discover candidate libraries separately for each parse tree in the forest according to the following steps:
\begin{enumerate}
    \item[S1.] We query the root node of the parse tree in the KG, which refers to a top-level module. 
    If there are module entities with the same name in the KG, we directly go to S2 with the query results. 
    Otherwise, we attempt to install a Python package with the same name as the top-level module.

    \item[S2.] There may be different versions or even different packages that have the same top-level module, so we filter these candidates further by submodules. 
    We set max-hop as the depth of the parse tree and query our KG to obtain spanning trees reachable from the top-level modules following the \textit{has\_module} relationships to max-hop.
    Then, we calculate the matching degree between the imported modules and the modules with \textit{import\_status} as \textit{True} in each spanning tree.
    The candidate spanning trees with the highest matching degree are retained.
    
    \item[S3.]
    Since the version updates may cause the addition or removal of attributes, we need to select the proper versions. 
    We enrich the spanning trees by querying the attributes of modules in the parse tree. 
    Similarly, we calculate the matching degree between the called attributes and the resources in each spanning tree. 
    The candidate trees with the highest matching degree are retained, and their corresponding libraries are optimal for the parse tree.
\end{enumerate}

After our discovery process, we finally determine the candidate libraries of each parse tree in the forest, which may correspond to multiple versions of more than one package. 
Meanwhile, the Python version with the maximum matching degree is selected. Python 3 is the default if their matching degrees are equal.

\subsection{Dependency Solving} 
\label{sec_solving}

To cover most third-party resources used in the target code, each top-level module should have at least one candidate library in the inferred runtime environment. 
Furthermore, we expect each package in the inferred environment to meet all its version constraints, which is called \emph{dependency solving}. 

Dependency solving is a hard problem in all non-trivial component models and has been proved NP-complete \cite{abate2020dependency,mancinelli2006managing}. 
Existing work treats dependency solving as a separate concern in package manager \cite{abate2012dependency} and relies on generic dependency solvers based on the tried-and-tested techniques such as solvers of the Boolean satisfiability problem (SAT) \cite{le2008sat}. 
There are efficient modern SAT solvers \cite{soos2009extending,EenS03,MahajanFM04}, but it is nearly impossible to control the answers provided by the solvers with many priorities.
For optional versions of a Python package, we prefer the latest version that can be installed successfully, since the latest version is usually downward compatible with previous versions and contains new resources.
For time efficiency and compliance with version selection priorities, we propose an efficient heuristic graph traversal algorithm. 

We create a start node, which points to all the virtual nodes that represent the top-level modules. We query our KGs for all transitive dependencies of the candidate libraries, forming a heterogeneous directed graph called a \emph{dependency graph}. The dependency graph would be iteratively extended until there are no more dependencies.
The nodes in the dependency graph are divided into conjunction nodes and disjunction nodes. 
Conjunction nodes include the start node and version nodes, which depend on all its direct successors.
Disjunction nodes include module nodes and package nodes, which require at least one direct successor.
Moreover, each edge pointing to a package node is attached with specific dependency requirements, which restrict the choice of its versions. 
The versions that fail to be installed during knowledge acquisition are not considered. 

\begin{definition}[Compatible dependency subgraph]
Let $G = (V, E)$ be a dependency graph, where $V$ denotes the vertex set including packages, versions and modules, and $E$ denotes the directed edge set based on the dependency relationships.
A compatible dependency subgraph $G' = (V', E')$ of $G$ satisfies:
\begin{enumerate}
    \item[(\romannumeral1)] $V' \subseteq V, E' \subseteq E$, and start node $s \in V'$.
    \item[(\romannumeral2)] We denote edge $a \to b$ by $(a, b)$, and regard a specific version requirement as a collection of versions that satisfy the requirement. Compatibility requires:

\begin{itemize}
    \item For each conjunction node $c \in V'$, $\forall\, e=(c, n) \in E, e \in E'$.
    \item For each disjunction node $d \in V'$, let $S = \{e\,|\,e=(d, n) \in E\}, |E' \cap S| \ge 1$ ($\,= 1$, if $d$ is a package node).
    \item For each package node $p \in V'$, let $R = \{\text{requirement of}$ $e\,|\,e=(i, p) \in E'\}$, 
    $\exists\, (p, n) \in E', n \in \bigcap_{r \in R}r$.
\end{itemize}
\end{enumerate}
\end{definition}

\smallskip
\noindent\textbf{SAT solver.} With the dependency graph, each node corresponds to a Boolean variable and the constraints are encoded as a Boolean formula in conjunctive normal form (CNF). 
The constraints of a dependency graph are translated into Boolean clauses in CNF according to the following rules:
\begin{itemize}
    \item Start node $s$ is set to \textit{True}, denoted by $(s)$.
    \item A conjunction node $c$ decides all its direct successors $x_1,x_2,$ $\dots , x_n$: $c \to (x_1 \wedge \dots \wedge x_n) \equiv \bigwedge_{1 \leq i \leq n} (\neg c \vee x_i)$. Conjunction nodes include the start node and version nodes.
    \item A disjunction node $d$ decides at least one of its direct successors ${x_1, x_2, \dots , x_n}$: $d \to (x_1 \vee \dots \vee x_n) \equiv (\bigvee_{1 \leq i \leq n} (x_i) \vee \neg d)$. Disjunction nodes include module nodes and package nodes.
    \item A virtual module node $m$ requires at least one of its candidate versions $v_1, v_2, \dots , v_n$: $m \to (v_1 \vee v_2 \vee \dots \vee v_n) \equiv (\bigvee_{1 \leq i \leq n} (v_i) \vee \neg m)$. Meanwhile, each candidate version $v_i$ requires its package $p_i$: $\bigwedge_{1 \leq i \leq n} (\neg v_i \vee p_i)$.
    \item A package node only has one of its versions $v_1, v_2, \dots , v_n$: $\bigwedge_{1 \leq i < n, i < j \leq n} (\neg v_i \vee \neg v_j)$.
    \item A version node $v$ is incompatible with versions $v_1, v_2, \dots , v_n$ of its direct dependencies that do not meet the specific requirements: $v \to (\neg v_1 \wedge \neg v_2 \wedge \dots \wedge \neg v_n) \equiv \bigwedge_{1 \leq i \leq n} (\neg v \vee \neg v_i)$.
\end{itemize}

If the formula is satisfiable, the resulting compatible dependency subgraph would consist of all the nodes with value \textit{True} and the edges between them. 

\begin{algorithm}[!t]
    \caption{$ExtendsSubgraph(G',n,p)$}
    \label{alg:dp}
    \KwIn{subgraph $G'$ to be extended, current traversed node $n$, direct predecessor $p$ of $n$ in this visit}
    \KwOut{$True$ or $False$ (indicating whether a compatible subgraph can be found.)}
    
    $tmpG \leftarrow G'$, and add node $n$ and edge $(p, n)$ to $tmpG$\;
    \If{$n$ is a conjunction node}
    {
        \ForEach{direct successor $c$ of $n$}
        {
            \If{$\neg ExtendsSubgraph(tmpG, c, n)$}
            {
                \Return $False$\;
            }
        }
        $G' \leftarrow tmpG$; \Return $True$\;
    }
    \Else
    {
        $C \leftarrow$ version-sorted direct successors of $n$ which are compatible in $tmpG$\;
        \If{exists edge $(n, d) \in tmpG$}
        {
            \lIf{$d \in C$}{$G' \leftarrow tmpG$; \Return $True$}
            \lElse{remove $(n, d)$, and recursively remove the dependencies of version $d$ from $G'$}
        }
        \While{$C$ is not empty}
        {
            \If{$ExtendsSubgraph(tmpG, C[0], n)$}
            {
                $G' \leftarrow tmpG$; \Return $True$\;
            }
            \lElse{remove $C[0]$ and the elements that have the identical dependency requirements in $C$}
        }
        \Return $False$\;
    }
\end{algorithm}

\smallskip
\noindent\textbf{Our heuristic algorithm.} The details of our algorithm are shown in Algorithm \ref{alg:dp} and we start traversing from the start node with an empty subgraph. 
With the depth-first search (DFS), our algorithm has two heuristic strategies: 
(\romannumeral1) Priority: we prefer the latest version (Line 13) that can be successfully installed by sorting all compatible versions (Line 8); 
(\romannumeral2) Pruning: if one version of a package is incompatible, we skip the other versions of that package with identical dependency requirements (Line 15), because they would cause the same conflicts. 
When the currently selected version of a package does not satisfy the newly added version constraints, the current version and all its direct dependencies and transitive dependencies need to be removed from the subgraph (Line 11). 
We only backtrack on the current search path when we encounter a dependency conflict, which is highly efficient but in rare cases may miss solutions. 
To address it, we call a SAT solver as the fallback to ensure the completeness of our algorithm when it claims that there is no solution.

\begin{figure}[tb]
  \centering
    \includegraphics[width=0.99\linewidth]{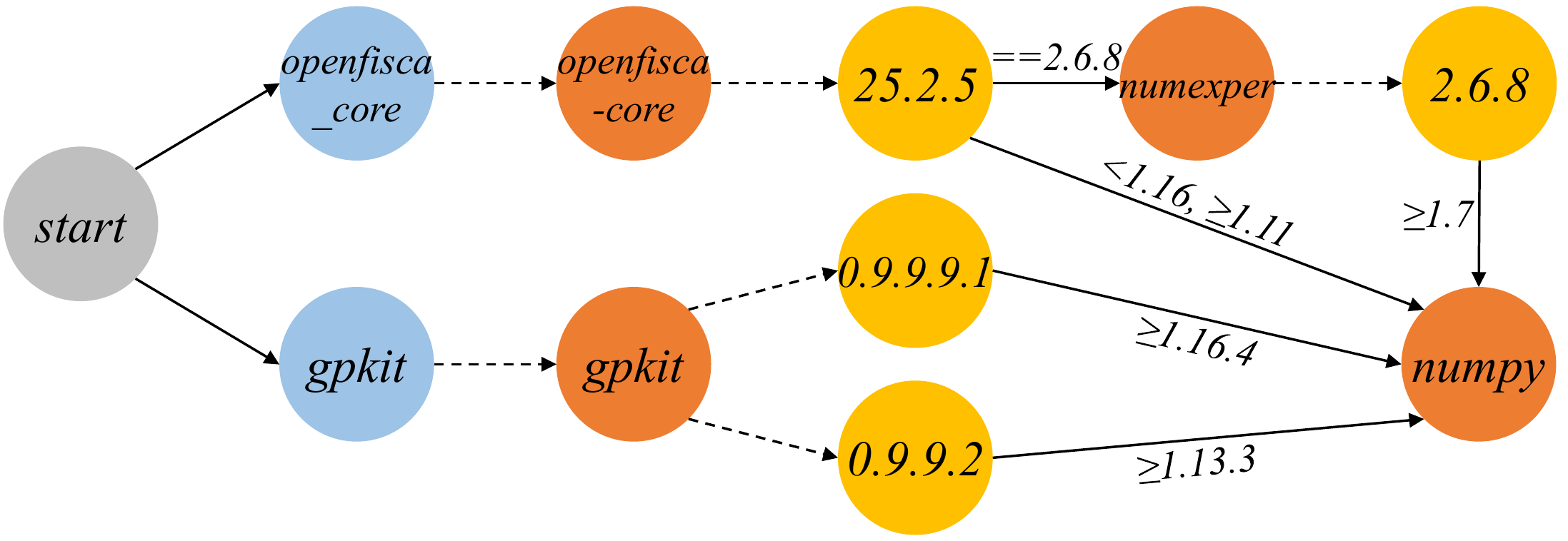}
  \caption{Partial display of a dependency graph. Node colors are in accord with the colors of entity types in Figure~\ref{fig_kg} except for the start node. Nodes with dashed outgoing edges are conjunction nodes, and nodes with solid outgoing edges are disjunction nodes. Best viewed in color.}
  \label{fig:reqGraph}
\end{figure}

Let us see the dependency graph in Figure~\ref{fig:reqGraph}.  
According to DFS, our algorithm first traverses to \textit{openfisca-core-25.2.5}, depending on \textit{numexper-2.6.8} that requires \textit{numpy$\ge$1.7}. 
Thus, the latest version \textit{1.16.6} is chosen for \textit{numpy}. 
When considering the direct dependency of \textit{openfisca-core-25.2.5} on \textit{numpy}, it is found that \textit{numpy-1.16.6} does not meet the new version requirements \textit{$<$1.16,$\ge$1.11}.
So, \textit{numpy-1.16.6} is removed (it does not depend on other packages) and a compatible version \textit{1.15.4} is reselected. 
When traversing to the package \textit{gpkit}, we prefer the latest version \textit{0.9.9.9.1}.
However, its version requirement \textit{$\ge$ 1.16.4} for \textit{numpy} conflicts with the previous requirement \textit{$<$1.16,$\ge$1.11}, so we reselect the version of \textit{gpkit}. 
We skip version \textit{0.9.9.9}, which has the identical requirement for \textit{numpy} as the incompatible version \textit{0.9.9.9.1}, and traverse to version \textit{0.9.9.2}. 
The version requirements of \textit{gpkit-0.9.9.2} are compatible and the current version of \textit{numpy} meets all the requirements. 
Now, we find a compatible dependency subgraph using our heuristic algorithm. 

At last, we identify the packages that need to be explicitly installed as well as the installation order. 
After removing the start node and module nodes in the subgraph, all packages with an in-degree of 0 are required to be installed explicitly, since they are not the dependencies of any packages. 
Moreover, for a package, if the latest version that meets its version constraints is different from the selected version, it would probably lead to an unexpected version, so we install this package explicitly as well. 
The installation order of all packages is generated by topological sorting, which ensures that all dependencies of each package are installed ahead of time.

\section{Evaluation}
\label{sec:exp}

\subsection{Experiment Settings}
\noindent\textbf{Dataset.} We conduct our experiments on the Gistable \cite{horton2018gistable} dataset, which is a real-world dataset built on the GitHub's gist system and contains 10,250 single-file Python code snippets. DockerizeMe and V2 have also been evaluated on this dataset.

\smallskip
\noindent\textbf{Comparative methods.}
In addition to our PyCRE, several comparative methods are assessed in the experiments: 
\begin{itemize}
\item Gistable \cite{horton2018gistable}, which attempts to install the Python packages with the same names as the imported top-level modules. It is a baseline method and does not use any knowledge. 
\item DockerizeMe \cite{horton2019dockerizeme}, which infers the runtime environments based on a pre-built knowledge base, without considering Python versions nor specific dependency versions. 
\item V2 \cite{horton2019v2}, which reuses the inference results of DockerzieMe as the starting environments in Python 2 and Python 3, and guides the version changes by the error messages of code execution until finding a working environment. 
\item SnifferDog \cite{wang2021restoring}, which uses a pre-built API bank to restore the execution environments of Jupyter notebooks. 
\end{itemize}

\smallskip
\noindent\textbf{Prior knowledge.} For the fairness of our experiments, all methods should acquire knowledge for the same Python packages. 
We obtain a list of Python packages from the knowledge base built by DockerizeMe \cite{horton2019dockerizeme}. 
According to Python Enhancement Proposals (PEP) 508~\cite{pep508}, Python package names are case-insensitive and do not distinguish between dash, dot and underscore. 
So, we merge the duplicate packages such as \textit{Flask-RESTful} and \textit{flask\_restful}. 
Eventually, we have 10,765 unique Python packages for knowledge acquisition.

The knowledge base built by DockerizeMe for the 10,765 Python packages is available. 
V2 uses the inference results of DockerzieMe as the starting environments, so it can be ignored. 
It is worth mentioning that the knowledge acquisition of DockerizeMe is not fully automated but requires manual work.
SnifferDog only acquires knowledge for a special selection of 488 Python packages. In this paper, we build its API bank for the 10,765 packages using its publicly available code. 
We also automatically construct our KGs for Python 2 and Python 3 by the method described in Section~\ref{sec:ka}. 
The knowledge of PyCRE and other methods is stored in a Neo4j database, except for SnifferDog, which is loaded directly into memory according to its design.

\smallskip
\noindent\textbf{Validation environment.} We leverage Docker to ensure that each Python code executes in an isolated environment. 
For Python 2 and Python 3, our Docker images are initialized on the official images \textit{python:2.7.18} and \textit{python:3.8.11}, respectively. 
We update pip to the latest version, currently 20.3.4 and 21.2.4 for Python 2 and Python 3, respectively.
We configure pip to use its new dependency resolver by default, which refuses to install the packages with incompatible requirement combinations. 
We conduct all experiments on Ubuntu 18.04 LTS with one Intel Xeon Gold 5117 CPU and 16GB memory.

\smallskip
\noindent\textbf{Experiment procedure.} 
Given that only our PyCRE and V2 can infer the Python version for target code and V2 cannot specify a fixed Python version, we conduct the experiments in three environments: Python 2, Python 3 and the inferred Python versions.

For each method, we first infer dependencies for each code in the dataset and record the time spent on the inference to analyze the efficiency. 
The timeout of inference is set to one hour, since V2 is a dynamic inference method, which needs to repeatedly roll back the versions and generate the validation environments. 
The code to infer dependencies in SnifferDog is incomplete, so we replicate it according to their paper for enabling SnifferDog to infer the dependencies of Python code. 

Then, we record the inferred Python dependencies in \textit{requirements.txt} in order and build the Docker image according to the experimental settings. 
To avoid the failures due to network fluctuation or other factors, we give ten minutes for each Python package installation when building each Docker image for validation. 
Failures to install dependencies when building images are ignored, which is in line with the behavior of programmers.

Finally, we run the code in the corresponding Docker container. 
The code snippets that run successfully are marked as \texttt{Success}.
We mark the code snippets running for more than one minute as \texttt{Timeout}, because some code may not stop running due to awaiting input, encountering dead loops, etc. 
The others are marked as \texttt{Exception} and the names of their exceptions are recorded. 
Since V2 only returns dependencies when it finds a working environment, for each failed code, we select the exception name of the last validation from its output logs.
Built-in exception \texttt{ImportError} is raised when an \texttt{import} statement fails to find the module definition or when a \texttt{from ... import} statement fails to find the name to import. 
We also include Python 3's built-in exception \texttt{ModuleNotFoundError} as \texttt{ImportError}, which is a subclass of \texttt{ImportError} and raised by \texttt{import} when a module cannot be located. 

\begin{table}[tb]
\centering
\caption{Statistics of the exit status of code execution on the Gistable dataset.}
\label{tab:dataset}
\begin{tabular}{lccc}
\toprule
            Exit status & Python 2 & Python 3 \\ 
\midrule    
            \texttt{Success}            & 2,112  & 1,293  \\
            \texttt{Timeout}            & \ \, 132   & \ \ \ \, 71    \\
            \texttt{ImportError}        & 5,515  & 3,536  \\ 
            \texttt{SyntaxError}        & \ \, 719   & 4,388  \\
            Other Exceptions   & 1,772  & 3,212  \\
\bottomrule
\end{tabular}
\end{table}

\smallskip
\noindent\textbf{Dataset analysis.} 
We run each code in the Gistable dataset in the validation environment and the exit status is shown in Table~\ref{tab:dataset}. 
There are 4,388 code snippets with \texttt{SyntaxError} in Python 3, much more than 719 in Python 2, which include 454 code snippets that have syntax errors in both versions. 
The difference in the number of code snippets that run successfully also indicates the bias towards Python 2 in the dataset. 
All automatic inference methods only focus on dependencies, i.e. \texttt{ImportError}, because it is almost impossible to automatically resolve errors in the code itself or missing external inputs such as database \cite{mondal2019can}. So, our experiments only validate the inferred environments of code snippets with \texttt{ImportError}.

\subsection{Evaluation of Knowledge Graph}

\smallskip
\noindent\textbf{Our knowledge graphs.} 
Knowledge acquisition is indeed time-consuming, which takes about 135 hours using 20 normal CPU cores. 
The rest of the building process is negligible in comparison. 
The whole process can be easily accelerated by parallelization, thus reducing time and effort to a great extent.
Table~\ref{tab:KGAnalysis} shows the scale of our Python package KGs and the quantities of each type of entities and relationships. 
Excluding those packages for which no version is available in the corresponding Python environments, we analyze 10,623 and 10,564 packages in Python 2.7.18 and Python 3.8.11, respectively. 
The extra 2,325 and 2,375 packages are added due to package dependencies, but we do not perform knowledge acquisition for these packages to ensure the fairness of knowledge. 
By further analyzing our KGs, we find that 74,657 (25.6\%) and 68,281 (22.2\%) versions fail to be installed, and 2,337,425 (20.7\%) and 2,360,264 (21.0\%) modules fail to be imported in Python 2 and Python 3, respectively, due to various issues such as missing dependencies and incompatible Python versions. 
This indicates that the releases on PyPI cannot always be installed successfully, and the modules and attributes obtained by statically parsing the code cannot always be used successfully.
Moreover, nearly 74\% of the successfully installed versions have at least one direct dependency, showing the necessity to consider the compatibility of inferred dependencies. 

\begin{table}[tb]
    \centering
    \caption{Statistics of our Python package KGs.}
    \label{tab:KGAnalysis}
    \begin{subtable}[t]{0.46\linewidth}
    \centering
    \resizebox{\textwidth}{!}{
    \begin{tabular}{lrr}
        \toprule
         Entity & Python 2 & Python 3 \\
         \midrule
         Package & 12,948 & 13,439 \\
         Version & 291,805 & 307,841 \\
         Module & 11,316,114 & 11,240,182 \\
         Attribute & 772,774 & 853,389 \\
         \midrule
         Total & 12,393,641 & 12,414,851 \\
         \bottomrule
    \end{tabular}}
    \label{tab:entities}
    \end{subtable}
    \begin{subtable}[t]{0.53\linewidth}
    \centering
    \resizebox{\textwidth}{!}{
    \begin{tabular}{lrr}
        \toprule
        Relationship & Python 2 & Python 3 \\
        \midrule
        \textit{has\_version} & 291,805 & 307,841 \\
        \textit{has\_module} & 11,316,114 & 11,240,182 \\
        \textit{has\_attribute} & 125,554,883 & 124,215,617 \\
        \textit{requires} & 717,357 & 946,092 \\
        \midrule
        Total & 137,880,159 & 136,709,732 \\
        \bottomrule
    \end{tabular}}
    \label{tab:rels}
    \end{subtable}
\end{table}

\smallskip
\noindent\textbf{Domain knowledge.} 
To compare the knowledge acquired by different methods, we count the quantities of versions, modules and attributes, which are shown in Table~\ref{tab:KGComparison}. 
The modules and attributes with the same full-qualified names are counted only once, 
and the packages have at least one version. 
DockerizeMe analyzes the latest version of each package to get the name of top-level modules, containing the fewest domain knowledge. 
SnifferDog downloads all releases on PyPI for each package and statically parses the Python code to get all the defined APIs. 
It fails to get any top-level modules for 4,831 packages, mainly because it cannot infer the Python versions needed for static parsing. 
In contrast to our PyCRE, SnifferDog gets slightly more versions and attributes, but cannot determine the availability of this knowledge, which may cause unexpected errors in the inferred environments. 
Additionally, we add submodules into the Python package ontology. 
The acquisition of submodules provides a large amount of module knowledge for PyCRE, which are useful to infer appropriate dependencies.

\smallskip
\noindent\textbf{Knowledge coverage on dataset.} 
We analyze the coverage of the knowledge acquired by each method on the Gistable dataset, which significantly influences the results of inference. 
Considering that the dataset is more biased towards Python 2, we parse each code under Python 2.7.18 by the approach presented in Section~\ref{sec_parse}. 
We divide the imported modules into top-level modules and submodules, since only our method supports queries for submodules. 
The analytical results are shown in Table~\ref{tab:coverage}. 
Compared to DockerizeMe and SnifferDog, our KG has the maximum coverage on all types of knowledge, but still lacks a large amount of knowledge about the dataset. 
Only 800 (46.5\%) top-level modules are covered in our KG, which means that in many cases PyCRE can only choose to install the Python package with the same name. 
Submodules and attributes are covered with 1,622 (22.9\%) and 3,486 (40.1\%), respectively. 
It increases the difficulty of discovering candidate libraries, which reflects the realistic limitations of exact matching and the necessity of our matching degree. 


\begin{table}[tb]
    \centering
    \caption{Comparison of domain knowledge acquired by different methods.}
    \label{tab:KGComparison}
    \begin{tabular}{lrrrr}
         \toprule
         Method & Package & Version & Module & Attribute \\
         \midrule
         DockerizeMe & 10,441 & \ \ 11,254 & \ \ \ \ 9,517 & - \\
         SnifferDog & \textbf{10,638} & \textbf{316,376} & \ \ \ \ 5,764 & \textbf{4,580,920} \\
         PyCRE (Python 2) & 10,623 & 291,805 & \textbf{338,709 }& 4,343,530 \\
         PyCRE (Python 3) & 10,564 & 307,841 & 302,908 & 3,949,917 \\
         \bottomrule
    \end{tabular}
\end{table}

\begin{table}[tb]
    \centering
    \caption{Comparison of knowledge coverage on the Gistable dataset.}
    \label{tab:coverage}
    \begin{tabular}{lccc}
         \toprule
         Method & Top-level module & Submodule & Attribute \\
         \midrule
         Dataset    & 1,721 & \,7,083 & \,8,704 \\
         DockerizeMe & \ \ \,727 & - & - \\
         SnifferDog & \ \ \,513 & - & \,1,924 \\
         PyCRE (Python 2) & \ \ \textbf{800} & \textbf{1,622} & \textbf{3,486} \\
         \bottomrule
    \end{tabular}
\end{table}

\subsection{Evaluation of Inference}
We evaluate the effectiveness and efficiency of PyCRE in inferring Python compatible runtime environments. 
Table~\ref{tab:res} shows the validation results of the inferred environments by each method.

\begin{table}[tb]
    \centering
    \caption{Validation of the inferred environments generated by each method in different Python releases.}
    \label{tab:res}
    \begin{subtable}[t]{0.99\linewidth}
    \centering
    \begin{tabular}{lcccc}
         \toprule
         Method & \texttt{ImportError} & \texttt{Success} & \texttt{Timeout} & \texttt{Others}\\
         \midrule
         Dataset & 5,515 & 2,112 & 132 & 2,491 \\
         Gistable & 2,592 & 2,988 & 422 & 4,248 \\
         DockerizeMe & 2,624 & 2,986 &415 & 4,225 \\
         SnifferDog & 2,296 & 3,086 & 466 & 4,402 \\
         \textbf{PyCRE} & \textbf{1,645} & \textbf{3,309} & \textbf{499} & \textbf{4,797} \\
         \bottomrule
    \end{tabular}
    \caption{Environment validation in Python 2.}
    \label{tab:res_py2}
    \end{subtable}
    \begin{subtable}[t]{0.99\linewidth}
    \centering
    \begin{tabular}{lcccc}
         \toprule
         Method & \texttt{ImportError} & \texttt{Success} & \texttt{Timeout} & \texttt{Others}\\
         \midrule
         Dataset & 3,536 & 1,293 & \ \ 71 & 5,350 \\
         Gistable & 1,751 & 1,934 & 218 & 6,347 \\
         DockerizeMe & 1,965 & 1,903 & 183 & 6,199 \\
         SnifferDog & 1,632 & 1,960 & 254 & 6,404 \\
         \textbf{PyCRE} & \textbf{1,302} & \textbf{2,114} & \textbf{270} & \textbf{6,564} \\
         \bottomrule
    \end{tabular}
    \caption{Environment validation in Python 3.}
    \label{tab:res_py3}
    \end{subtable}
    \begin{subtable}[t]{0.99\linewidth}
    \centering
    \begin{tabular}{lcccc}
         \toprule
         Method & \texttt{ImportError} & \texttt{Success} & \texttt{Timeout} & \texttt{Others}\\
         \midrule
         V2 & 2,654 & 3,073 & 379 & 4,144\\
         \textbf{PyCRE} & \textbf{1,524} & \textbf{3,410} & \textbf{579} & \textbf{4,737}\\
         \bottomrule
    \end{tabular}
    \caption{Environment validation with the inferred Python versions.}
    \label{tab:res_total}
    \end{subtable}
\end{table}

\smallskip
\noindent\textbf{\texttt{ImportError}.} 
The most intuitive assessment is the ability to resolve \texttt{ImportError}, which reflects the effectiveness of the runtime environment inference. 
As shown in Table~\ref{tab:res}, PyCRE resolves the most \texttt{ImportError} in all three different settings and is significantly better than the comparative methods. 

Gistable, the baseline method that is consistent with the programmer's behaviors in solving dependency issues, fails to resolve \texttt{ImportError} for 2,592 gists and 1,751 gists in Python 2 and Python 3, respectively. 
DockerizeMe fails to resolve \texttt{ImportError} in more gists than the baseline Gistable's method, mainly because of its method of matching target code and Python dependencies. 
DockerizeMe uses partial matching for imported modules to find libraries in the knowledge base, which is based on the longest prefix of the module name. 
For example, statement \textit{from pyspark.sql.functions import udf} in the code snippet\footnote{\url{https://gist.github.com/samuelsmal/feb86d4bdd9a658c122a706f26ba7e1e}} is mapped to the package \textit{py}, because there is no module \textit{pyspark} and the longest matched prefix is module \textit{py} in Dockerizeme's knowledge base.
However, the Python package corresponding to this module is \textit{pyspark}, which is the package that other methods choose to install. 
In fact, DockerizeMe uses domain knowledge to solve 482 code for which baseline cannot solve \texttt{ImportError}, but fails to solve 514 code for which baseline can solve \texttt{ImportError} in Python 2. 
Although these two methods do not specify versions for the inferred packages, which minimizes the version restrictions, they still encounter dozens of dependency conflicts.

Since V2 uses the inference results of DockerzieMe as the starting environments, its performance is similarly affected. 
V2 claims to find working environments for 3,206 code, but 133 of them do not run successfully in our validation, and even have dependency issues. 
One major reason is that V2 does not consider the compatibility between the packages in the inferred environments. 
Additionally, DockerizeMe and V2 install some packages incidentally according to their association rules, which leads to many redundant packages and is also more likely to cause dependency conflicts. 

SnifferDog achieves relatively good performance, mainly because it has a large amount of API knowledge to assist its inference. 
However, SnifferDog is designed for Jupyter notebooks and ignores some issues in Python, such as the inference of Python versions, which affects its applicability on more general Python code.

\smallskip
\noindent\textbf{Working environments.}
Another valuable metric is the number of working environments, which indicates the ability of an inference method to restore the runtime environments of Python code. 
As shown in Table~\ref{tab:res}, PyCRE infers a successfully runnable environment for the most Python code snippets in all three different settings. 
Moreover, the code marked as \texttt{Timeout} usually has a dead loop or awaits inputs, and can be treated as \texttt{Success} at least until they time out. In this sense, it can be assumed that PyCRE infers the working environments for 3,989 (38.9\%) gists, which is also the best result among all methods.


\smallskip
\noindent\textbf{Inference time.}
Inference time is a crucial metric for the efficiency of inference methods and represents the user's waiting time, which is one of the most important performance metrics for software. 
As listed in Table~\ref{tab:time}, we evaluate the average inference time and the longest inference time for each method. 
We only consider the 5,655 gists that have at least one third-party package in Python 2 or Python 3 and have no inference timeout in all methods. 
Since the API bank of SnifferDog is loaded in memory and does not query an external database, we do not discuss its inference time. 
The inference time of DockerizeMe and PyCRE, which fully uses pre-built knowledge bases, are short and comparable, 
whereas V2, which involves online execution, has a much longer inference time. It is worth noting that only the inference of V2 for 353 gists does not finish in one hour.

\begin{table}[tb]
    \centering
    \caption{Average and longest inference time of each method for 5,655 code snippets with at least one third-party package and have no inference timeout in all methods.}
    \label{tab:time}
     \begin{tabular}{lccc}
        \toprule
          Method &  DockerizeMe & V2 & PyCRE \\
          \midrule
          Avg. time (s) & \ \ \ \textbf{5.0} & \ \ 128.0 & \ \ \ \ 7.0 \\
          Max. time (s) & \textbf{120.6} & 3,570.7 & 215.2 \\
          \bottomrule
    \end{tabular}
\end{table}

\smallskip
\noindent\textbf{Ablation experiment.} 
To validate the effectiveness of our proposed heuristic algorithm, we exclusively use the SAT solver to conduct an ablation experiment (i.e. disable our heuristic algorithm), which is called PyCRE (SAT only). 
In the inferred Python version, PyCRE (SAT only) infers 3,337 environments marked as \texttt{Success} and 536 marked as \texttt{Timeout}, leaving 1,597 \texttt{ImportError}. 
PyCRE with the heuristic algorithm, shown in Table~\ref{tab:res_total}, is superior to PyCRE (SAT only) in the validation of inferred results. 
The average solving time of the heuristic algorithm is 0.2 seconds while the SAT solver is 3.9 seconds. 
However, the longest time of the heuristic algorithm is 8.4 seconds, which is much lower than 219.3 seconds of the SAT solver, and the difference becomes more significant as the size of the dependency graph increases.
Moreover, compared to our heuristic algorithm, only using the SAT solver has problems caused by randomness.
The SAT solver may choose old versions as long as the dependency requirements are met, but it usually leads to a loss of compatibility with newer versions. 
Also, the environments inferred by the SAT solver are variable and have the potential to introduce redundant dependencies, which may cause troubles for users. 
Based on our analysis, our heuristic algorithm solves 5,602
(99.4\%) of the 5,637 code snippets for which have at least one dependency and a compatible solution. We believe the approximation of our heuristic algorithm is good. 
Besides, while specifying a version for each inferred package can better match the target code, such version requirements are more likely to cause dependency conflicts. 
In contrast, our approach is effectively conflict-aware through dependency solving, with dependency conflicts occurring only in 18 inferred environments.

\begin{figure}[!t]
    \centering
    \includegraphics[width=0.9\linewidth]{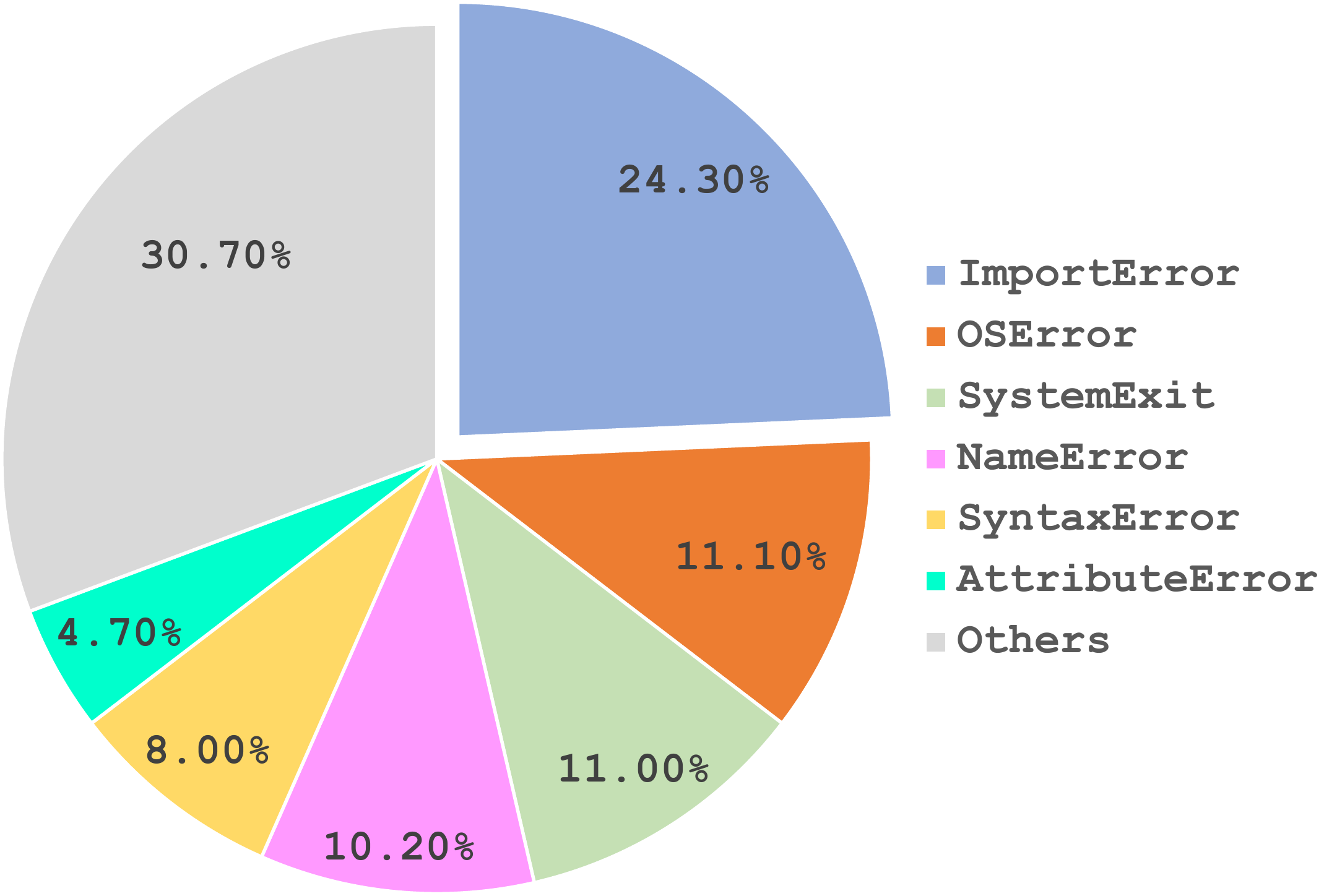}
    \caption{Proportions of 6,261 (\,$=$\,1,524\,$+$\,4,737) exceptions thrown by the environment validation inferred by PyCRE.}
    \label{fig:ExcepDistribution}
\end{figure}

\smallskip
\noindent\textbf{Exception statistics.}
We further analyze the exceptions in Table~\ref{tab:res_total} for the validated environments inferred by PyCRE.
We include the subclasses of an exception in the count of that exception and the results are shown in Figure~\ref{fig:ExcepDistribution}.
Except for \texttt{ImportError}, the three most common Python built-in exceptions are \texttt{OSError}, \texttt{SystemExit} and \texttt{NameError}, which are mainly caused by code design or missing extra inputs, and cannot be solved by installing dependencies. 
\texttt{SyntaxError} is raised largely due to the target code itself.
Moreover, \texttt{AttributeError} is raised when an attribute reference fails, mainly due to the reference on \texttt{NoneType} and some third-party resource.
Therefore, we can assume that in most cases, the packages are also compatible in the inferred environments where \texttt{ImportError} is resolved by PyCRE. 

\smallskip
\noindent\textbf{Practical significance.}
The effectiveness and efficiency of PyCRE contributes to software engineers solving dependency issues of Python code in practice. 
Beginners often have difficulty building a runtime environment for sample code, and professional programmers also waste much time with complicated versions and dependencies. 
As in the manual evaluation on the Gistable dataset, it takes software engineers between 20 minutes and 2 hours to build the runtime environment \cite{horton2018gistable}, while the average inference time of our approach on this dataset is only 7 seconds, which can significantly improve the efficiency of development. 
Considering the performance of PyCRE, it is also important for the development of automated configuration management.


\subsection{Threats to Validity}
The results of our experiments may suffer from several threats to validity. The first is the fairness of knowledge, which can largely affect the inferred results of individual methods. 
For this reason, all methods in our experiments acquire knowledge based on the same 10,765 packages following their own methods. 
However, the knowledge base shared by DockerizeMe and V2 was previously constructed, and the latest versions of Python packages in it are sometimes not the latest versions at now. 
We think that this has little impact on the experimental results because the dataset was proposed earlier than the construction of the DockerizeMe knowledge base, which means that the code actually relies on older versions of the packages. 
Oppositely, more new versions in PyCRE also increase the difficulty of inference.
Another threat is \texttt{Timeout} in the validation results. 
The code that is marked as \texttt{Timeout} does not throw any exceptions in a minute, but this does no guarantee that there are no problems in subsequent execution. 
Considering that the import statements for resources are usually at the beginning of the Python code, code that executes with a timeout usually does not have dependency issues, which mitigates this threat.

\subsection{Case Study}
In addition to resolving the compatibility of inferred environments, the following Python code snippets illustrate several other unique capabilities of PyCRE. 

\smallskip
\noindent\textbf{Skip unusable releases.}
PyCRE obtains the real status of package installations and module imports with knowledge acquisition, which can guide PyCRE to skip the releases that are actually unusable. 
See the code snippet excerpted from a real-world gist:\footnote{\url{https://gist.github.com/miratcan/4cd70e9515ab722b2bce}}

\begin{figure}[ht]
  \flushleft
  \includegraphics[width=0.8\linewidth]{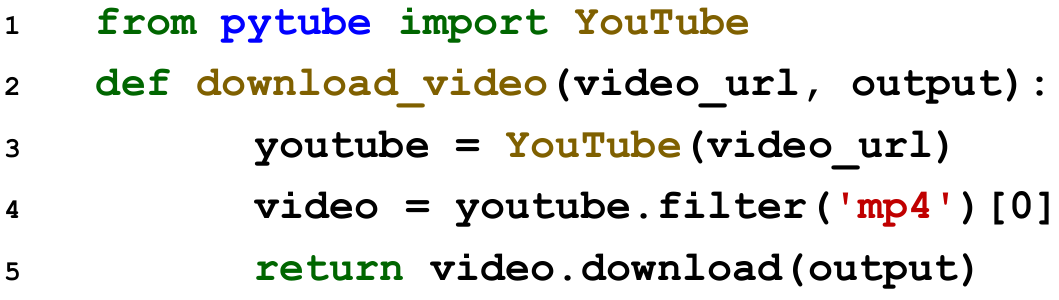}
\end{figure}


Note that this code is only compatible with Python 2. 
Gistable and DockerizeMe generate \texttt{pip install pytube} to install the dependency, which means that the latest version \textit{pytube-9.6.0} in Python 2 would be installed by default. 
However, after successfully installing the package, it fails to import \textit{pytube} in the code, which raises an exception due to the problem in this version itself. 
V2 fails to find a working environment for this code, because it encounters timeout while verifying the candidate environment, resulting in no information available to guide the version changes. 
SnifferDog statically parses all versions of \textit{pytube} and the latest version \textit{11.0.0} contains the attribute \textit{pytube.YouTube}, so it selects \textit{pytube-11.0.0}, which fails to be installed and is actually only available in Python 3. 
In the knowledge acquisition phase, we already know that the module cannot be successfully imported in these versions, so based on our pre-built KGs, PyCRE skips the unusable versions and installs \textit{pytube==9.5.2}.

\smallskip
\noindent\textbf{Avoid useless downloads.}
If one version of a package fails to be installed or is incompatible with other packages, pip would attempt to install other versions instead, which is also known as the backtracking behavior.\footnote{\url{https://pip.pypa.io/en/stable/user_guide/\#dependency-resolution-backtracking}}
Since pip does not have full package dependency information before downloading the package, it may lead to a large number of unnecessary downloads, which increases the time and system memory spent on building the runtime environment.
Skipping unusable releases as described above can avoid the useless downloads caused by failed installations. 
The \texttt{import} statements below exemplify the strength of our approach in another aspect. 

\begin{figure}[ht]
  \flushleft
  \includegraphics[width=0.8\linewidth]{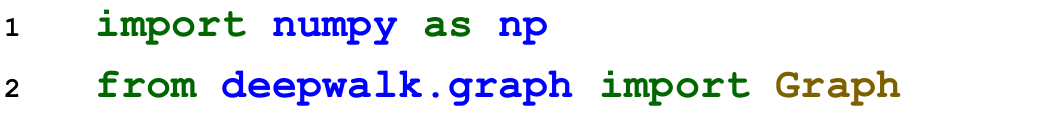}
\end{figure}


The dependencies that are inferred to be explicitly installed by all the methods except PyCRE are \textit{numpy} and \textit{deepwalk}. 
Pip first downloads \textit{numpy-1.16.6}, and then downloads \textit{deepwalk-1.0.3}.
However, \textit{deepwalk-1.0.3} depends on \textit{gensim-3.8.3}, which requires \textit{numpy$\le$1.16.1} and is incompatible with \textit{numpy-1.16.6}, so \textit{numpy-1.16.1} is downloaded again. 
This causes a redundant download for \textit{numpy}. 
Based on the unique knowledge of dependencies, PyCRE infers that only \textit{deepwalk-1.0.3} needs to be installed explicitly and that compatible \textit{numpy-1.16.1} would be installed automatically.

\section{Related Work} 
\label{sec:rel}

\subsection{Python Runtime Environment Inference}
There are many studies related to software KG in software engineering \cite{lin2017intelligent, wang2017construct, ren2020api, wang2019searching}.
DockerizeMe \cite{horton2019dockerizeme} is a pioneering work, which offline builds a knowledge base to infer the language/system-level environment dependencies required for Python code to execute without import errors. Compared with it, our PyCRE has several significant differences. First, DockerizeMe only considers the latest version of each package, which cannot handle removal or renaming that may accompany with version changes. 
Second, DockerizeMe maps the top-level modules to Python packages, and does not obtain submodules and attributes. However, there may be multiple versions of different packages containing modules with the same name, which needs further decision based on the attributes called in the code. 

V2 \cite{horton2019v2} enhances DockerizeMe by exploring the possible configuration space for a Python code snippet. 
It validates candidate environments iteratively through code execution and applies environment mutation to generate new candidate configurations according to the failure messages. While V2 can find successful runtime environments for some Python code, its feedback-directed search is quite time-consuming, and even fails when an incorrect version does not manifest as a crash.  On the contrary, PyCRE infers the appropriate Python libraries using the pre-built KGs, which is an efficient approach.

SnifferDog \cite{wang2021restoring} is committed to restoring the execution environments of Jupyter notebooks and even reproducing the results. 
It builds an API bank to record mappings from popular Python libraries to their APIs by parsing the Python files. 
Several aspects affect the effectiveness of its environment inference. First, its API bank stores only public functions, but other public submodules and variables should also be considered. 
Second, the static analysis may fail to get any knowledge due to the uncertainty of Python versions used by the releases.
Third, it does not guarantee that the defined APIs can actually be called, since the packages and modules where these APIs are located may fail to be installed or imported.

As far as we know, there is no study addressing the compatibility of the inferred environments like our work. 

\subsection{Dependency Solving}
Dependency solving (and some of its variants) has been proved to be NP-complete, which can be easily encoded into a SAT solving problem using CNF \cite{mancinelli2006managing,abate2012dependency,le2008sat}. Any solution of SAT is equally valid, but practically some solutions are better than others for dependency solving. Trezentos et al. \cite{trezentos2010apt,trezentos2010comparison} defined the software dependency problem as an extension of the SAT formulation called pseudo-Boolean optimization (PBO). There are several efficient PBO solvers, such as Open-WBO solver \cite{martins2014open} and sat4j solver \cite{le2010sat4j}, and pseudo-Boolean constraints can be translated into clauses that can be handled by a standard SAT solver \cite{een2006translating}. 

Abate et al. \cite{abate2020dependency} reviewed proposals from the dependency solving field in recent years. They treat dependency solving as a separate concern in component evolution management \cite{abate2012dependency}. Although a few popular package managers like Eclipse P2 use SAT solvers for dependency solving, the vast majority of package managers including pip still uses customized dependency graph traversals. 
The traditional dependency resolver of package managers receives a specific installation request given by the user, whereas dependency solving in PyCRE needs to determine the required installations, which is more difficult. 
Fortunately, we have global knowledge through the pre-built Python package KGs, which enables us to heuristically prune the search path for generating the compatible environments.

\section{Conclusion and Future Work}
\label{sec:conclusion}

In this paper, we propose an approach to automatically inferring Python compatible runtime environments with domain KG. 
We design a domain-specific ontology for Python third-party packages and propose an automatic approach to constructing the Python package KGs.
Given a Python code, we discover candidate libraries by measuring the matching degree with third-party resources used in the code. 
Furthermore, we propose a heuristic graph traversal algorithm to infer the compatible runtime environment.
Compared with existing approaches, we show the superior effectiveness, efficiency and compatibility of our approach in runtime environment inference. 
Our approach can contribute to automated software configuration management and facilitate code reuse. 

In future work, we will acquire knowledge for more Python packages and improve the coverage. 
We also plan to add the deprecation information of modules and APIs into the KGs, and use it to further infer appropriate versions.
Besides, we will extend the dependency inference to the entire project instead of single-file code and consider the compatibility with local dependencies, which is more general in practice.
Finally, we will apply our approach to other languages with transitive dependencies such as Node.js.


\balance
\section*{Acknowledgments}
We thank the anonymous reviewers for their valuable comments. This work is supported by the National Natural Science Foundation of China (No. 61872172), and the Collaborative Innovation Center of Novel Software Technology and Industrialization.


\bibliographystyle{ACM-Reference-Format}
\bibliography{ref}

\end{document}